\documentclass[conference,a4paper]{IEEEtran}
\usepackage[utf8]{inputenc}
\usepackage[english]{babel}

\usepackage{cite}
\usepackage{amsmath,amssymb,amsfonts}

\usepackage[ruled, lined, linesnumbered, commentsnumbered, longend]{algorithm2e}
\usepackage{algpseudocode}

\usepackage{graphicx}
\usepackage{textcomp}
\usepackage{xcolor}
\usepackage{soul}
\usepackage{enumerate}
\usepackage{comment}

\usepackage{multirow}
\usepackage{tabu}
\usepackage{float}
\usepackage[inline]{enumitem}
\usepackage{amsmath}
\usepackage{mathtools}

\usepackage[autostyle]{csquotes}
\usepackage[utf8]{inputenc}

\usepackage{textcase}

\usepackage[left=1.62cm,right=1.62cm,top=1.9cm]{geometry}

\definecolor{boristext}{rgb}{0.3, 0.36, 0.88}
\definecolor{boriscomments}{rgb}{0.83, 0.0, 0.0}
\definecolor{davidcomments}{rgb}{0.0, 0.0, 0.83}

\newcommand{\bb}[1]{\textcolor{boristext}{#1}} % for text added
\newcommand{\bcom}[1]{\textcolor{boriscomments}{[#1]}} % for comments

\DeclareMathOperator*{\E}{\mathbb{E}}  

\DeclareMathOperator*{\argmax}{arg\,max}

\usepackage{bbm}

\begin{document}

% This code is to reduce the list of authors by using et. al:
\bstctlcite{IEEEexample:BSTcontrol}

\title{Spatial Reuse in IEEE 802.11bn Coordinated Multi-AP WLANs: A Throughput Analysis}
%\title{A Throughput Model for Spatial Reuse in IEEE 802.11bn Coordinated Multi-AP WLANs}

\author{
\IEEEauthorblockN{David Nunez$^{\flat}$, Francesc Wilhelmi$^{\star}$, Lorenzo Galati-Giordano$^{\star}$,\\ Giovanni Geraci$^{\sharp}$\,$^{\flat}$, and Boris Bellalta$^{\flat}$\vspace{0.1cm}}
\IEEEauthorblockA{$^{\flat}$\emph{Department of Information and Communications Technologies, Universitat Pompeu Fabra, Barcelona, Spain}}
\IEEEauthorblockA{$^{\star}$\emph{Radio Systems Research, Nokia Bell Labs, Stuttgart, Germany}}
\IEEEauthorblockA{$^{\sharp}$\emph{Telef\'{o}nica Research, Barcelona, Spain}}
\IEEEauthorblockN{\thanks{Corresponding author: \emph{david.nunez@upf.edu}.}
}}
\maketitle

% Letter?
% https://pimrc2024.ieee-pimrc.org/ (February 2024)

% ---------------------------
% ---------------------------
% ---------------------------
% ---------------------------

\begin{abstract}
IEEE 802.11 networks continuously adapt to meet the stringent requirements of emerging applications like cloud gaming, eXtended Reality~(XR), and video streaming services, which require high throughput, low latency, and high reliability. To address these challenges, Coordinated Spatial Reuse (C-SR) can potentially contribute to optimizing spectrum resource utilization. This mechanism is expected to enable a higher number of simultaneous transmissions, thereby boosting spectral efficiency in dense environments and increasing the overall network performance. In this paper, we focus on the performance analysis of C-SR in Wi-Fi~8 networks. In particular, we consider an implementation of C-SR where channel access and inter-Access Point (AP) communication are performed over the air using the Distributed Coordination Function (DCF).
% \st{, and the specific group of devices that share the spatial resources is triggered based on the station selected by the transmit opportunity holder}. 
For such a purpose, we leverage the well-known Bianchi's throughput model and extend it to support multi-AP transmissions via C-SR. Numerical results in a WLAN network that consists of four APs show C-SR throughput gains ranging from 73\% to 284\% depending on the inter-AP distance and the position of the stations in the area. 
%\\\hl{Gio: rethink the title. I understand we cannot use the same title as Francesc (\emph{``Throughput Analysis of IEEE 802.11bn Coordinated Spatial Reuse"}), but I would only minimally change it. I doubt the red keywords are useful, and other keywords are missing like: \emph{Coordinated Spatial Reuse, Wi-Fi 8, IEEE 802.11bn, Multi-AP Coordination}.}
\end{abstract}

%~\cite{bianchi2000performance}

\begin{IEEEkeywords}
Coordinated Spatial Reuse, IEEE 802.11bn, Multi-Access Point Coordination, Wi-Fi 8. %\hl{Gio: No need.}%, Managed WLANs. 
\end{IEEEkeywords}

% ---------------------------------------------------
% ---------------------------------------------------
% ---------------------------------------------------
% ---------------------------------------------------
\section{Introduction}

Next-generation applications such as eXtended Reality~(XR), high-quality holographic video, and zero-delay file exchange to support cooperative and remote working require further improvements on wireless networks ~\cite{cavalcanti2020wireless}. For example, the throughput requirements of Virtual Reality (VR) applications must be above 400 Mbps to properly deliver a 4K VR 360º video experience~\cite{vodafone_VR_paper} and even higher (above 1 Gbps) to provide an ideal (extreme) VR experience~\cite{WBA_2023_report}. Technologies operating on it like Wi-Fi will be very important for those new applications because they consume a lot of spectrum resources \cite{OUGHTON2024102766}. Moreover, achieving those requirements is especially challenging in current dense scenarios with multiple Wi-Fi networks sharing the same channel, where the heightened contention and collision levels not only limit the achievable throughput but also impact the latency and reliability required by such stringent applications \cite{real_time_app_TIG_report_mentor_2009r6, latencyrequirementsEdgePaper,CarGerKni2024}.

To address the challenges imposed by next-generation applications in high-density Wireless Local Area Network (WLAN) scenarios, the future IEEE 802.11bn amendment \cite{UPF_Nokia_wifi8,UHRobjectives}  is likely to introduce Multi-Access Point Coordination (MAPC). This framework is envisioned to enhance the overall latency and reliability performance, through alleviating channel access contentions in dense Wi-Fi deployments. 
%MAPC facilitates the coordination of transmissions among overlapping APs, thus potentially enhancing the way large Wi-Fi networks are managed and operated. 

%Enabling MAPC offers various benefits, including contention mitigation between APs and their associated stations, and improving the utilization of spectrum resources for a more efficient spatial reuse as well. 
MAPC enables the introduction of different methods that perform coordinated resource allocation, such as Coordinated Time Division Multiple Access (C-TDMA), where Access Points (APs) agree on splitting time resources, or Coordinated Orthogonal Frequency-Division Multiple Access (C-OFDMA), where the agreement is on the allocated Resource Units (RUs) \cite{UPF_Nokia_wifi8}. In certain scenarios, higher performance can also be achieved by enabling simultaneous transmissions through Coordinated Spatial Reuse (C-SR) and more complex techniques like Coordinated Beamforming (C-BF) or Joint Transmission (J-TX). Among these techniques, C-SR stands out as a particularly promising scheme since it can reduce latency and, in most scenarios, increase throughput, without requiring complex and resource-consuming measurement acquisition and coordination signaling \cite{CSRoverheads_mentor_Sony_0616r0, CSRoverheads_mentor_Qualcomm_0120r0, CSRoverheads_mentor_Offino_0095r0}. For this reason, C-SR stands out as one strong candidate to be included in the next 802.11bn amendment.

\begin{figure*}[t!!!!!!!!!!!!!!!!!!!!]
    \centering
    \includegraphics[scale=0.60]{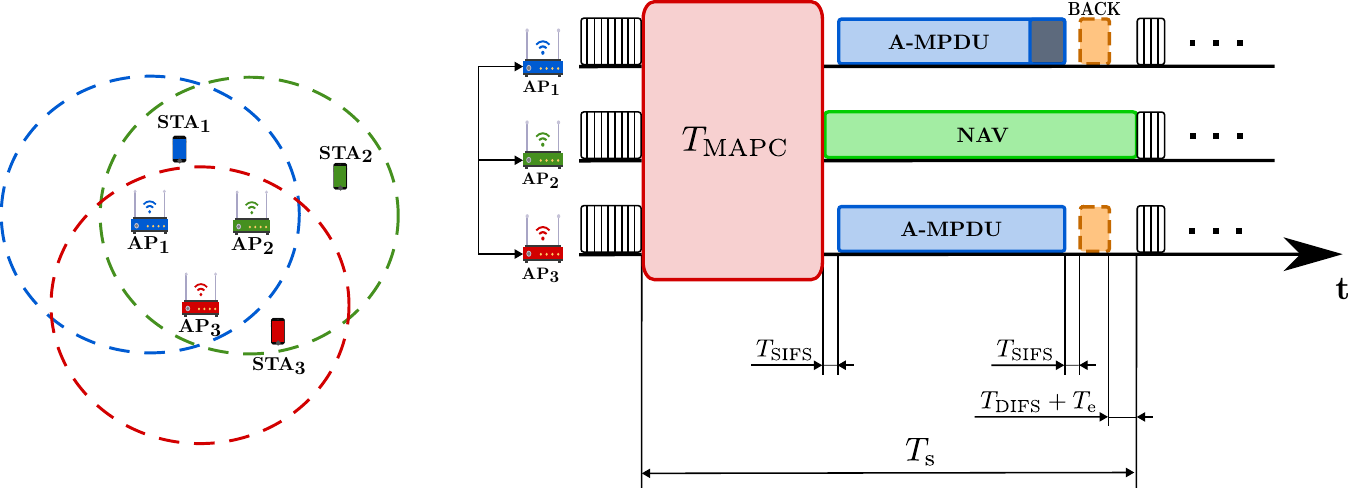}
    \caption{Left side: Example of the proposed MAPC operation in a small OBSS deployment. Right side: coordination frames and coordinated transmission, where AP$_{1}$ and AP$_{3}$ are selected due to their spatial reuse compatibility.}
    \label{Fig:timeline}
\end{figure*}

The performance of MAPC techniques is still not fully understood. The definition of the MAPC framework started with IEEE 802.11be~\cite{AdrianGarciaSurvey}, but was then postponed until IEEE 802.11bn. Since then, multiple flavors of MAPC have been proposed, some advocating for decentralized solutions \cite{MentorSR1534r1} and others for central controllers \cite{MentorCoordinationCentr0665}. Another relevant point of discussion is related to the way coordination signaling should be performed, and whether over wireless or wired connections \cite{MentorSR1534r1, MentorCoordinationFiber1186}. 
%Alternative proposals include a multi-AP and multi-robot coordination framework for optimized resource utilization (as a solution for autonomous edge networks in Industry 4.0)~\cite{davecavalcanti_Lnorm_algorithm_industry4}, or a multi-AP coordination system architecture enabling Deep Reinforcement Learning (DRL)-based channel access~\cite{coordinationAccessDeepRL}. 
Close in spirit to our work, \cite{mio_MAPC_groups_scheduling} introduced a framework where compatible multi-AP groups are scheduled for performing simultaneous transmissions within a TDMA-slotted Transmission Opportunity (TXOP). The results in~\cite{mio_MAPC_groups_scheduling} evidenced that algorithms based on per-AP selection are preferable to those based on per-group selection.

When it comes to C-SR, its performance has been evaluated in a few IEEE 802.11 standard contributions \cite{MentorSR1534r1,MentorSR0107r1,MentorSR0576r1} via simulation results, showing the potential gains compared with legacy access, C-TDMA, and C-OFDMA. In \cite{TXOPsharingPaper}, the authors evaluated the throughput of networks implementing C-TDMA and C-TDMA with SR (C-TDMA/SR), showing that C-TDMA/SR leads to throughput gains between 50\% and 140\%, depending on the scenario, when compared to C-TDMA. As for the analytical characterization of C-SR, a model based on Markov chains was introduced in \cite{francesc_wifi8_markov} to compute the C-SR's throughput and its spatial efficiency % and showed a considerable gain of 59\% for the former, 
in comparison to the Distributed Coordination Function (DCF). However, the model in~\cite{francesc_wifi8_markov} did not consider backoff collisions and coordination overheads, which may affect the achievable performance gains. We aim to overcome such limitations.

%thus hindering the actual gains achieved by the mechanism.
%C-TDMA alone was also evaluated in~\cite{stefano_cTDMA} using simulations, which showed a network latency reduction of up to one order of magnitude with respect to legacy channel access.

%In this work, we contribute to the development and evaluation of MAPC features by studying the gains of C-SR. More specifically, our contributions are as follows:
Our contributions can be summarized as follows:
\begin{itemize}
    \item We introduce a C-SR mechanism relying on the existence of spatial reuse groups of AP-Station (STA) pairs, i.e., groups of devices that can transmit simultaneously. Spatial reuse groups are periodically updated (e.g., every few hundred msecs) based on the exchanged Received Signal Strength Indicator (RSSI) reports between APs. Then, once the spatial reuse groups are updated, the data exchange phase starts, where TXOP sharing is used by the channel access winner to signal group transmissions.
    %We assume that all AP-STA pairs know to which groups they belong, so when an AP-STA pair is selected (i.e., the AP that has won the contention selects a given STA for a transmission), the group of AP-STA pairs compatible with that AP-STA pair is signaled. 
    %
    \item We leverage Bianchi's model to analyze the C-SR mechanism and evaluate its potential performance under full-buffer conditions. The presented analysis accommodates different group creation and selection algorithms, thus enabling to model and compare different implementations.
    %\bcom{General model given the group tx probs are known --> focus on test different group creation strategies}.
    %Our modeling accounts for the fact that each group is selected with a probability proportional to the number of AP-STA pairs it contains to keep the analysis simple.
    %
    \item Using the proposed analytical model, we evaluate the throughput of C-SR and compare it against DCF. Our results illustrate how the achievable throughput is affected by the devices' location. Moreover, they demonstrate the gains achieved by the proposed group formation approach, especially in scenarios with a large number of STAs per AP.
    %the trade-off between the number of devices per group and the amount of interference allowed.
    % achieving gains of up to \hl{$200\%$} for SR-friendly topologies.
\end{itemize}

This paper focuses on throughput to evaluate the gains by leveraging C-SR capabilities compared to the traditional DCF in Wi-Fi networks. These gains can be turned into latency improvements and reliability, which is crucial for supporting applications with stringent performance requirements.

%\bcom{Paragraph selling the paper, comparing against the state of the art. Focus on the analysis.}

%\bb{0) Insist on collisions and backoff; i) To identify where the multi-ap coordination gains appears i.e., (6); ii) High Perf. gains for any (reasonably) small distance, in more than > 20 \% of the scenarios; iii) The model is general, and it may allow us to optimize different parameters, such as: CW, Ts, test different group creation algorithms}

%\red{The rest of the paper is organized as follows. Section~\ref{section:CSR} presents the considered C-SR operation and a method to create groups of AP-STA pairs that transmit simultaneously. In Section~\ref{section:throughput_analysis}, we present the proposed throughput analysis method for characterizing IEEE 802.11bn networks implementing C-SR. In Section~\ref{section:performance_eval}, we define the scenario and the system model along with numerical results for evaluating the proposed C-SR mechanism against the legacy DCF operation. Finally, Section~\ref{section:conclusions} concludes the paper.}%
%
%\footnote{\hl{Gio: Not needed for such a short paper.}}
%

% ---------------------------------------------------
% ---------------------------------------------------
% ---------------------------------------------------
% ---------------------------------------------------

\section{Proposed C-SR Mechanism Model}\label{section:CSR}
% ---------------------------------------------------
% ---------------------------------------------------

% ----------------------------------------------------
% ----------------------------------------------------
\subsection{C-SR Operation}

We make the following modeling assumptions for the sake of tractability:
%To study the achievable throughput of C-SR, we make some considerations to keep the system model tractable and avoid implementation-side details, which are left outside the goals of this work:
\begin{itemize}
    \item We consider a group of APs within the same coverage area, i.e., Overlapping Basic Service Set (OBSS), that contend to access the channel using the default Carrier-Sense Multiple Access with Collision Avoidance (CSMA/CA) mechanism. 
    \item The C-SR groups, i.e., the groups of AP-STA pairs that can benefit from SR concurrent transmissions, are computed based on the RSSI values exchanged between the coordinated APs. For low mobility scenarios, the RSSI varies on a slow time scale and its acquisition incurs limited overhead.
    %Before C-SR transmissions happen, and based on information exchanged between all coordinated APs (i.e., RSSI values), the groups of AP-STA pairs that can benefit from SR concurrent transmissions are computed. This procedure can be periodically repeated to account for changes in the network.  %computed beforehand. %\bcom{before C-SR transmissions happen? The beforehand sounds as if it was not targetted in this paper}\dcom{true, it does not clearly specify when,  but I couldn't find a way to state this properly. Any ideas?}
    %
    \item The C-SR operation is implemented atop DCF in a TXOP-sharing fashion, i.e., when a coordinated AP gains access to the channel (i.e., the backoff counter reaches zero) it initiates a multi-AP transmission along with compatible transmitters in the OBSS. The AP that wins the contention (leader AP) then selects one of its STAs as a target receiver, and triggers the group (if exists) that includes other AP-STA pairs compatible with the target receiver. 
\end{itemize}

The implementation of the overall coordination approach is illustrated in Fig.~\ref{Fig:timeline}, where it can be seen how a set of coordinated APs compete to gain access to the channel using DCF. To capture the coordination overheads, we assume a coordination phase of duration $T_{\rm MAPC}$. This phase includes channel reservation and information about the set of devices and settings to be used in the incoming coordinated TXOP, such as bandwidth, Modulation and Coding Scheme (MCS), TXOP duration, etc.%
\footnote{We assume that the coordination phase is feasible and aligned with the MAPC mechanism being discussed and likely included as part of the IEEE 802.11bn amendment. However, details about its implementation are out of the scope of this paper.}
After the $T_{\rm MAPC}$ period, AP$_{1}$ and AP$_{3}$ start transmitting data in Aggregate MAC Protocol Data Units (A-MPDUs) to their corresponding STAs, while AP$_{2}$ sets its Network Allocation Vector (NAV), waiting its turn for decrementing its backoff. Subsequently, Block ACKs (BACK) are sent to AP$_{1}$ and AP$_{3}$ by their corresponding STAs using OFDMA.

% Notice that while the computation of the groups is centralized, and done by the CC, the channel access and per-transmission signaling are done following the default 802.11 operation, thus guaranteeing a fair coexistence with external Wi-Fi networks, if exist. 

% ---------------------------------------------------
% ---------------------------------------------------
\subsection{Multi-AP Group Formation}\label{section:group_formation}

An SR group consists of several APs and STAs pairs that are compatible, i.e., pairs of devices that can transmit simultaneously without causing the Signal-to-Interference-plus-Noise Ratio (SINR) at the served STAs to fall below a threshold $\gamma_{\textrm{CE}}$ (see Section~\ref{section:performance_eval} for further details). The groups are defined based on the  RSSI perceived by each STA from the neighboring APs.\footnote{Owing to time division duplexing and channel reciprocity, downlink RSSI can be estimated from measurements on uplink frames (data or ACK frames).} 

% More specifically, all the possible combinations of AP-STA pairs $j\in \mathcal{M}$ that can transmit simultaneously are computed through an exhaustive search. \bcom{What is $\mathcal{M}$?}.

%\bcom{Although not needed here, before defining $\mathcal{M}$, I would introduce that we have $K$ APs, and each AP $j$ has $S_j$ stations associated. It may help to say that the size of $\mathcal{M}=|\mathcal{M}|=\sum_{j=1}^{M}{S_j}$. Otherwise, we reach the scheduling rates, and $S_j$ suddenly appears there.}\dcom{makes sense}\bcom{go on!} \dcom{Esto sigo sin tenerlo muy claro, al final estamos definiendo  AP$_j$ como uno de los $K$ APs, por lo que $j = 1, 2, ..., K$. Luego decimos que el indice $j$ hace referencia al AP-STA pair en combination $i$, que no necesariamente tendrá tamaño K.}

Consider $\mathcal{M}$ as the collection of all AP-STA pairs and $\mathcal{C}$ as the ensemble of all valid combinations attainable (via exhaustive search) from the pairs within $\mathcal{M}$. Then, we define $\mathcal{M}_i \subseteq \mathcal{M}$ as the subset of the $M_i = |\mathcal{M}_i|$ active AP-STA pairs in the ${i}$-th combination, $i\in \mathcal{C}$. Moreover, the quality of each combination $i$ is assessed by considering the total number of packets that can be transmitted by the $M_i$ APs in combination $i$, i.e.,  $\rho_i = M_i\sum_{\substack{\forall j \in \mathcal{M}_{i} \\ }}\varrho_{i,j}$, where $\varrho_{i,j}$ is the estimated number of packets correctly transmitted by AP-STA pair $j\in \mathcal{M}_{i}$, computed based on the MCS index employed by AP-STA pair $j$.
%
%

%i.e.,  $\rho(i) = M(i) \sum_{\substack{j=1 \\ }}^{M(i)}\rho_{j}(i)$, where $\rho_{j}(i)$ is the estimated number of packets correctly transmitted in AP-STA pair $j\in \mathcal{M}_{i}$, computed based on the Modulation Coding Scheme (MCS) index employed by AP-STA pair $j$}. 
% Note that by multiplying by $M'(i)$ each group we can fairly compare groups with a different number of AP-STA pairs, i.e., groups with more AP-STA pairs suffer more interference ---lower MCSs--- but have more chances to be scheduled.

Using the computed combinations of feasible simultaneous transmissions, a subset $\mathcal{G}\subseteq \mathcal{C}$ is selected. Furthermore, to guarantee that all the AP-STA pairs are granted with a minimum number of transmission opportunities, we establish a constraint whereby a given AP-STA pair $j$ can only be selected exactly $\varphi$ times in $\mathcal{G}$, i.e., $\sum_{i\in \mathcal{G}} \mathbbm{1}_{[j\in \mathcal{M}_i]} = \varphi$. For the sake of simplicity and without loss of generality, we set $\varphi=1$ for the remainder of this paper, which means that an AP-STA pair $j$ will only appear in a single group $i\in \mathcal{G}$.
%
%Moreover, assuming $R=1$ implies that the probability that group $i$ is scheduled is equivalent to the probability that any of the AP-STA pairs in group $i$ transmits.

To illustrate how the groups $\mathcal{G}$ are selected, we consider the scenario in Fig.~\ref{Fig:ToyScenarios}(a), where four BSSs (each formed by one AP and one STA) use the same channel. We assume that the RSSI measurements from all the AP-STA pairs in the scenario are available to every AP. Table~\ref{table:combinations} shows an excerpt of the combinations of AP-STA pairs and their associated score $\rho_i$. To choose among all possible groups, we sort them in descending order starting by the ones with the highest values of $\rho_i$. Moreover, the groups wherein one or more AP-STA pairs cannot decode frames properly due to the high interference generated are discarded, even if their associated $\rho_i$ value is high. 
% Based on the proposed group selection mechanism, the selection of AP-STA pairs for groups $\mathcal{G}$ is done by using the table, in a way that it is guaranteed that each AP-STA pair only appears $R$ times.
Following this mechanism, in the example shown in Table~\ref{table:combinations}, we would select combinations $\mathcal{G}=\{1,4,5\}$.

% Notice that selecting a higher number of pairs per group is not always the best option in terms of performance, as a result of the interference inflicted during the transmission.
% (e.g., $\rho(1) + \rho(12) +  \rho(13) >  \rho(11)$, i.e., $530 > 326$). 

% \begin{figure}[t!!!!!!!!!!!]
%     \centering
%     \includegraphics[scale=0.8]{scenario_5APs_5STAs.pdf}
%     \caption{Scenario with 5 BSSs, each formed by an AP and one STA.}
%     \label{Fig:scenario}
% \end{figure}

\begin{table}[h]
\centering
\caption{Combinations obtained for Deployment 1.} 
\begin{tabular}{|l|l|l|l|l|l|}
\hline
\textbf{}    & \textbf{AP$_1$} & \textbf{AP$_2$} & \textbf{AP$_3$} & \textbf{AP$_4$} &   $\rho_i$ \\ \hline
\textbf{1}  & STA$_{1}$            & -           & -            & STA$_{4}$               & 1740         \\ \hline
\textbf{2}  &  STA$_{1}$           & -            & -            & -             & 453       \\ \hline
\textbf{3} &      -        &    -          &     -         &      STA$_{4}$        &   453           \\ \hline
\textbf{4} & -            & STA$_{2}$            & -            & -              & 407         \\ \hline
\textbf{5} & -            &   -         &     STA$_{3}$        &     -          & 362          \\ \hline
\end{tabular}
\label{table:combinations}
\end{table}

% ---------------------------------------------------
% ---------------------------------------------------

\subsection{Group Transmission  Probability}\label{sec:scheduling}

Every time an AP accesses the channel, i.e., its back-off counter reaches zero, it randomly selects one of its STAs for transmission and, if the TXOP sharing is feasible based on the defined groups, it triggers the corresponding multi-AP group. Therefore, the probability that a group transmits is proportional to the number of AP-STA pairs that it contains. Thus, for group $i$, the transmission 
probability can be computed as:
\begin{align}
    \label{Eq:scheduling_rate}
%    \phi_i = \frac{1}{K}\sum_{\forall j\in \mathcal{M}_i}{\frac{1}{S_j}}\text{,}
    \phi_i = \sum_{\forall j\in \mathcal{M}_i}{\alpha_j}\text{,}
\end{align}
%where $K$ is the total number of APs in the network, and $S_{j}$ is the number of stations associated with the AP in the AP-STA pair $j$ of the coordination group.   
% where $K$ is the total number of APs in the WLAN, and 
where $\alpha_j$ is the probability that AP-STA pair $j\in \mathcal{M}_i$, and therefore group $i$, is selected for transmission (i.e., STA $j$ is selected for transmissions since its corresponding AP wins the channel contention). Following the full-buffer traffic assumption, and assuming that all the APs use the same access parameters and that STAs associated to each AP are selected with the same probability, we have that $\alpha_j=~\frac{1}{K S_j}$, where $K$ is the total number of APs in the network, and $S_j$ the number of STAs associated to AP $j$. %Similarly, given we consider an AP-STA pair $j$ can belong only to a single group, the probability that AP $j$ transmits is equivalent to the probability AP-STA pair $j$ is scheduled. Finally, $\sum_{\forall i \in \mathcal{C}}{\phi_i}=1$.

%Let us take a closer look into the expression (\ref{Eq:scheduling_rate}). Note that, ${1}/{K}$ is the probability that a given AP access the channel, which is the same for all APs since we assume full-buffer traffic and they use the same channel contention parameters. 

For instance, the transmission probability of the groups selected in the example of Section~\ref{section:group_formation}, i.e., $\mathcal{G}=\{1,4,5\}$, are calculated based
on the probabilities that the involved STAs are served at a given TXOP. Therefore, the transmission probability of group 1, $\phi_1 = \frac{1}{4}+\frac{1}{4} = 1/2$ (notice that AP$_{1}$ and AP$_{4}$ only have one associated STA). Additionally, $\phi_4 = \phi_5$ = 1/4, because AP$_{2}$ and AP$_{3}$ transmit alone in their corresponding groups and have only one associated STA.

\section{Throughput Analysis}\label{section:throughput_analysis}

To investigate the attainable full-buffer throughput for C-SR, we extend Bianchi's DCF model \cite{bianchi2000performance} by introducing the concept of multiple AP transmissions enabled by C-SR within a single TXOP. Bianchi's DCF model applies to overlapping sets of devices and assumes a slotted time system, where the minimum slot unit corresponds to an empty slot ($T_{\rm e}=9\mu$s). The empty slot is also used as the unit for decreasing the backoff counter by one slot while contending for the channel. Apart from the empty slot, we find successful and collision slots. In the original Bianchi's model, a successful slot contains the transmission of only one device. However, in the context of C-SR, it can contain multiple simultaneous transmissions from the AP-STA pairs of a C-SR group. A collision slot, on the other hand, occurs when two or more devices finish their backoff countdown at the same time. 

%Notice that the Bianchi model is used here to characterize the channel access that dictates the selection of the leader AP in MAPC.

%Apart from the empty slot, we find successful and collision slots in attempting to win the channel contention.The successful slot contains the transmission of only one device, and in the context of our model, it is able to trigger several AP-STA pairs to participate in the TXOP which results from the random computation of a unique backoff value by the TXOP winner. A collision slot, on the other hand, occurs when two or more devices finish their backoff countdown at the same time. Notice that the Bianchi model is used here to characterize the channel access that dictates the selection of the leader AP in MAPC.

To compute the throughput of a given AP-STA pair $j$, we begin by calculating the probability of a device accessing the channel in a random slot time, given by:
\begin{align}
    \tau = \frac{1}{\mathbb{E}[B] + 1}\text{,}
\end{align}
where $\mathbb{E}[B]$ is the expected backoff in number of slots. Considering an infinite number of retransmissions per packet and utilizing the Binary Exponential Backoff (BEB) \cite{BELLALTA_EB}, we can compute $\mathbb{E}[B]$ as:
\begin{align}
%    \mathbb{E}[B] = \frac{\mathrm{CW_{\rm {min}}}+1}{2}\left(\frac{1-p_{\rm {c}}-p_{\rm {c}}(2p_{\rm {c}})^{m}}{1-2p_{\rm {c}}}\right) - \frac{1}{2} \textrm{,}
\mathbb{E}[B] = \frac{\mathrm{CW_{\rm {min}}}+1}{2}\left(\frac{1-p-p(2p)^{m}}{1-2p}\right) - \frac{1}{2} \textrm{,}
\end{align}
where $\mathrm{CW_{\rm {min}}}$ denotes the minimum contention window, $m$ is the number of backoff stages, and $p=1-(1-\tau)^{K-1}$ represents the conditional collision probability. Since the conditional collision probability $p$ depends on $\tau$, and vice versa, both parameters need to be iteratively computed until convergence.

% Next, to compute the throughput achieved by the AP-STA pair $j$, \hl{we define $\phi_{i}$ as the scheduling rate of group $i$} where $j$ is contained \bcom{$\phi_i$ should be better defined in the system model}. 
%\bcom{As it is now, the formula includes the case in which an AP-STA pair can belong to different groups, right? If yes, we should mention it, as before we have said we were going to focus in the case that an AP-STA pair can belong to a single group.}\dcom{Yes, that's right. Actually it was in the previous versions of this text, but I wanted to remove it since we are making emphasis in the case of $R = 1$ for this paper. Talking about several groups for each STA could be confusing to the reader. It's probably worth it as a footnote?}\bcom{No me queda claro que quieres hacer aquí. Formula solo para R=1 y footnote para queue?} \dcom{No no, lo que digo es que no me gustaría mencionar nada sobre que cada STA puede pertenecer a various grupos, aún cuando la expresión es general y admite esto. Hemos hablado en este paper siempre de 1 grupo para cada STA, i.e.,  R = 1.} 
\begin{comment}
Thus, we define the throughput of AP-STA pair $j$ as follows:
\begin{align}\label{eq:per_STA_cSR_throughput}
\Gamma_j = \frac{p_s L \left(\sum_{i\in \mathcal{G}}{\mathbbm{1}_{[j\in \mathcal{M}_i]}\phi_{i} N_{i,j}}\right)}{\mathbb{E}[T]}\text{,}
\end{align}
\end{comment}

The average duration of a slot, $\E[T]$, can be expressed as:
\begin{align}\label{eq:slot_duration}
    \E[T]=p_{\rm {e}} T_{\rm {e}} + p_{\rm {s}} \sum_{\forall i\in \mathcal{G}}{\phi_{i} T_{{\rm {s}},{i}} + p_{\rm {c}} T_{\rm {c}}} \text{,}
\end{align}
where, $p_{\rm {e}}=(1-\tau)^{K}$ and $T_{\rm {e}}$ are the probability and the duration of an empty slot given the coexistence of $K$ transmitters, respectively, $p_{\rm {s}} = K\tau(1-\tau)^{K-1}$ represents the probability of a successful transmission slot, $T_{{\rm {s}},i}$ represents the duration of a transmission when group $i$ is selected, and $T_{\rm {c}}$ is the duration of a collision slot. Besides, the probability that a slot includes a collision can be computed as $p_{\rm {c}} = 1-p_{\rm {e}}-p_{\rm {s}}$. 

A successful transmission comprising group $i$ has a duration of $T_{{\rm {s}},i}$, as it considers the maximum duration among all AP-STA pairs that transmit simultaneously in group $i$, i.e., $\max_{j\in \mathcal{M}_i} \{T_{i,j}\}$, and each of them depends on its own selected transmission rate, $R_{i,j}$. Thus the duration for AP-STA $j$ can be computed as
\begin{align}
    T_{i,j} = &  T_{\rm MAPC} + 2 T_{\rm SIFS} + T_{\rm DATA}(R_{i,j}) \nonumber \\  
    & + T_{\rm BACK} + T_{\rm DIFS} + T_{\rm {e}} \text{,}
\end{align}
where all the signaling frames are transmitted using the basic rate.

Additionally, the aggregate throughput of the WLAN is
\begin{align}\label{eq:per_STA_cSR_throughput}
\Gamma =  \frac{p_{\rm {s}} L \bigg( \sum_{i \in \mathcal{G}} \phi_{i} \big( \sum_{j\in \mathcal{M}_i} N_{i,j} \big) \bigg)}{\mathbb{E}[T]},
\end{align}
where $L$ denotes the packet size, and $N_{i,j}$ is the A-MPDU size in number of packets that the STA in $j$ receives from its corresponding AP for a transmission in group $i$. The term $\sum_{i \in \mathcal{G}} \phi_{i} \big( \sum_{j\in \mathcal{M}_i} N_{i,j} \big) $ represents the average number of packets transmitted in a successful slot, and it characterizes the throughput gain of C-SR compared to DCF. Notice that~\eqref{eq:per_STA_cSR_throughput} holds for the particular case of DCF (without C-SR) if considering that each group contains a single AP-STA pair.

\begin{comment}
Finally, the throughput of AP-STA pair $j$ is
\begin{align}\label{eq:per_STA_cSR_throughput_individual}
\Gamma_j = \sum_{i\in \mathcal{G}}{\mathbbm{1}_{[j\in \mathcal{M}_i]}} \phi_{i} \Gamma.
\end{align}
\end{comment}

Finally, we can compute the throughput of AP-STA pair~$j$ by simply considering the number of transmitted packets by the target AP-STA pair in~\eqref{eq:per_STA_cSR_throughput} instead of the entire group, i.e.,     
\begin{align}\label{eq:per_STA_cSR_throughput_individual}
\Gamma_j =  \frac{p_{\rm {s}} L \bigg( \sum_{i \in \mathcal{G}} \phi_{i} \big( \mathbbm{1}_{[j\in \mathcal{M}_i]} N_{i,j} \big) \bigg)}{\mathbb{E}[T]}.
\end{align}

% Since the amount of data the STA receives depends on the specific group, we provide the expected performance of such a pair by averaging the throughput it obtains across all the groups in which it is included. 

%\bcom{To mention that in the analysis we consider groups are scheduled stochastically, but that follows the round robin strategy provided by the CC}

% ---------------------------------------------------
% ---------------------------------------------------
% ---------------------------------------------------
% ---------------------------------------------------

\section{Performance Evaluation}\label{section:performance_eval}

\subsection{Scenario and System Model}

%\bcom{Do we mention full-buffer traffic? We have said it at the beginning, but perhaps we need to repeat it here. Similarly, do we say that all APs have the same number of stations?}

We consider an OBSS with $K=4$ full-buffered APs deployed throughout a squared scenario (see deployments in Fig.~\ref{Fig:ToyScenarios}), with each AP separated by a minimum distance of $d_{\rm {AP-AP}}$ meters from the others. The APs contend to access the same channel using DCF (they are assumed to be within the communication range of the others), so packet collisions may occur during that process. Each AP $k$ has $S_{k}$ stations associated, which serves in a downlink manner. The STAs are randomly placed $d_{\rm {AP-STA}}$ meters away from their corresponding AP. For the sake of simplicity, we consider the same number of STAs associated with each AP.

The transmission rate employed in a given AP-STA transmission depends on the MCS employed (we consider 802.11ax MCS values), which is selected based on the estimated RSSI at the receiver. As a result, stations close (far) from their AP are served using a higher (lower) MCS. The maximum number of aggregated packets in each A-MPDU depends on the MCS. We consider that all the APs use the same fixed transmit power. For an SR transmission to be successful, it is required that the SINR at the receiver surpasses a predefined capture effect threshold $\gamma_{\textrm{CE}}$  \cite{lee2007CaptureEffect}. Notice, as well, that this condition strongly depends on the set of concurrent AP-STA pairs selected for transmission in a TXOP. 

The path loss effects $P_{\rm {L}}$(dB)  are modeled using the TGax model for Enterprise Scenarios
\cite{pathloss}:
\begin{equation}\label{pl_equation}
P_{\rm {L}} = 40.05 + 20\log_{10}\left(\frac{\min(d,B_{\rm {p}})f_{\rm {c}}}{2.4}\right) + \mathbbm{1}_{d>B_{\rm p}} P' + 7W_{\rm {n}} \text{,}
\end{equation}
where $d$ is the distance between the transmitter and the receiver in meters (but no less than 1), $f_{\rm {c}}$ is the carrier frequency in GHz, $W_{\rm {n}}$ is the number of walls traversed and $P'$ is given by $P' = 35\log_{10}(d/B_{\rm {p}})$. 
% when $d$ is higher than the breaking point $B_{p}$. Otherwise, it is zero.

%\subsection{Results}
We validate the results obtained with our analytical model using a simulator developed in Matlab reproducing IEEE 802.11 channel access in detail. Table~\ref{tab:simulation_paramters} collects the parameters used in the evaluation. Note that we consider the same $T_{{\rm {s}},i}$ for all groups' transmissions, denoted by  $T_{{\rm {s}}}$, which results in different $N_{i,j}$ values for each STA and group. 

% ---------------------------------------------------
% ---------------------------------------------------
\subsection{Example Deployments}
 \begin{figure*}[t!!!!!!!!!!]
    \centering
    \includegraphics[scale=0.50]{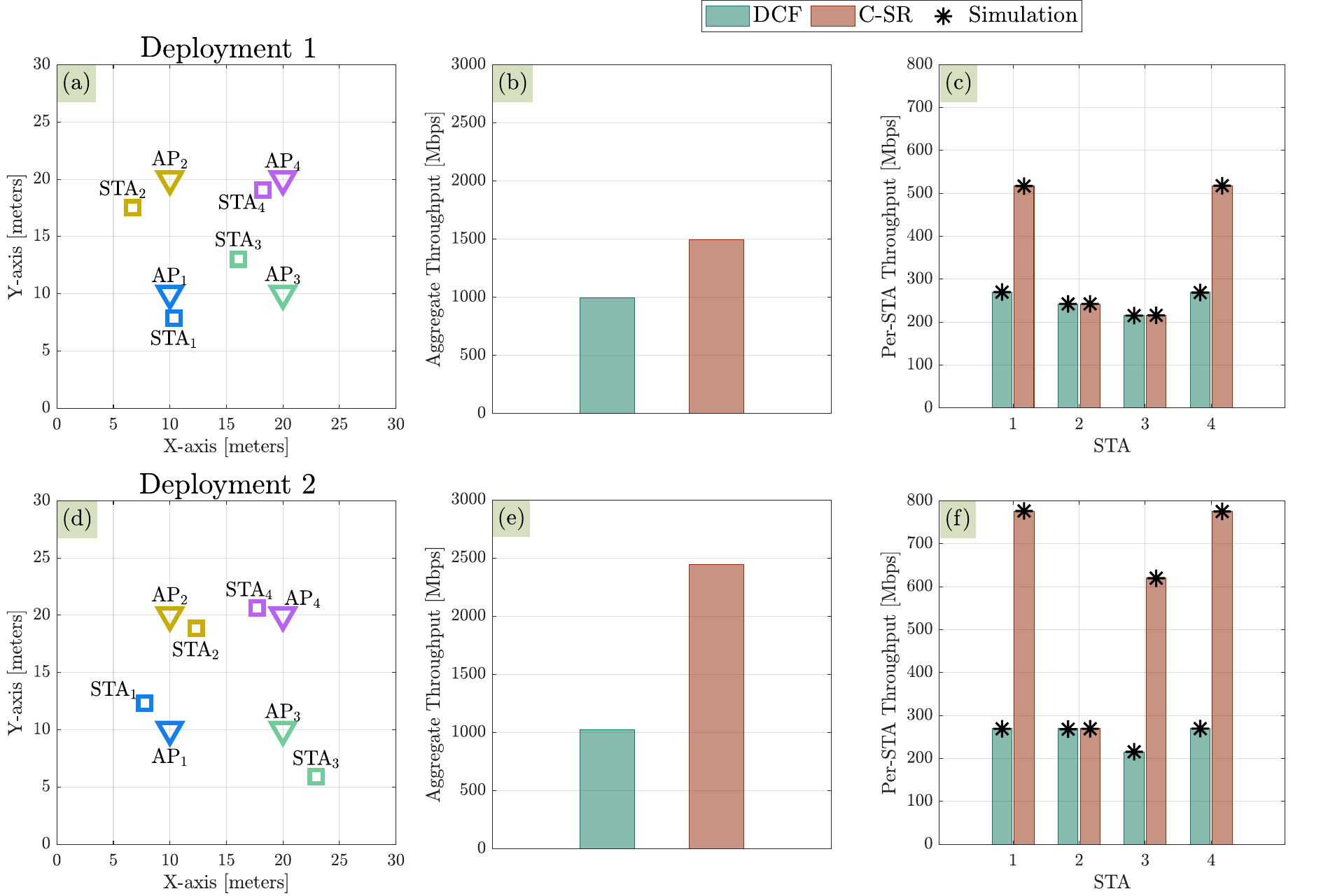}
    \caption{Deployment 1 and Deployment 2 are shown in (a) and (d), respectively. APs have been placed to a distance $d_{\rm {AP-AP}} = 10$ meters, and 1 STA is associated to each of them. The aggregate throughput of DCF and C-SR in each deployment is shown in (b) and (e). Moreover, the throughput per STA in Deployment 1 and Deployment 2 are shown in (c) and (f), respectively.}
    \label{Fig:ToyScenarios}
\end{figure*}

% --------------------------------_
% -------------------------------_
% --------------------------------_
% --------------------------------_
The two %static
deployments presented in this section have been selected to illustrate how the performance of C-SR varies from one deployment to another, depending on the devices' location.  Fig.~\ref{Fig:ToyScenarios}(a) and Fig.~\ref{Fig:ToyScenarios}(d) show the topology of Deployment~1 and Deployment~2, respectively. In both cases, we have $K=4$ APs, the distance between access points is $d_{\rm {AP-AP}} = 10$ meters, and a single STA per AP is randomly deployed $d_{\rm {AP-STA}} \in [1, 10]$ meters away from it. Besides, Fig.~\ref{Fig:ToyScenarios}(b) and Fig.~\ref{Fig:ToyScenarios}(e) show the aggregate throughput (i.e., $\Gamma$) achieved under the DCF and C-SR schemes. Additionally, the bars drawn in Fig.~\ref{Fig:ToyScenarios}(c) and Fig.~\ref{Fig:ToyScenarios}(f) show the individual throughput experienced by each station in both deployments, when DCF and C-SR are applied. Black asterisks represent their corresponding throughput values obtained by simulation, showing a perfect match between analysis and simulations.

%Deployment 1 is an example of a non-favorable deployment for C-SR, where only AP$_{1}$ and AP$_{4}$ are suitable to transmit simultaneously to STA$_{1}$ and STA$_{4}$, respectively, while AP$_{2}$ and AP$_{3}$ have to transmit alone. As indicated in Table~\ref{table:combinations} the selected groups are $\mathcal{G}=\{1,5,6\}$ (see Section~\ref{section:group_formation}). The aggregate throughput in Fig.~\ref{Fig:ToyScenarios}(b) shows a gain of 51\% for C-SR with respect to DCF. This result is due to the higher number of gained TXOPs, and consequently, more transmissions by AP$_{1}$ and AP$_{4}$ because their scheduling rate is $\times2$ compared to the APs that transmit alone, i.e., $ \phi_{1} = 1/2$ and $\phi_{5} = \phi_{6} = 1/4$.  

Deployment 1 is an example of a non-favorable deployment for C-SR, where only STA$_{1}$ and STA$_{4}$ can be served simultaneously, while STA$_{2}$ and STA$_{3}$ require to be served using alone transmissions. As indicated in Table~\ref{table:combinations}, the selected groups in this case are $\mathcal{G}=\{1,4,5\}$ (see Section~\ref{section:group_formation}). The aggregate throughput in Fig.~\ref{Fig:ToyScenarios}(b) shows a gain of 50\% for C-SR with respect to DCF. This result is due to the higher number of gained TXOPs, and consequently, more transmissions by AP$_{1}$ and AP$_{4}$ because their group transmission probability is twice as high as the one of the APs that transmit alone, i.e., $ \phi_{1} = 1/2$ and $\phi_{4} = \phi_{5} = 1/4$. %\hl{What about $\phi_{4}$?} 

On the other hand, the STAs in Deployment 2 in Fig.~\ref{Fig:ToyScenarios}(d) are better located to form compatible C-SR groups. In this case, AP$_{1}$, AP$_{3}$, and AP$_{4}$ are selected to perform C-SR transmissions to their corresponding STAs, leaving only AP$_{2}$ alone, hence boosting the network's performance by around 139\% with respect to DCF in Fig.~\ref{Fig:ToyScenarios}(e). The per-STA throughput in Fig.~\ref{Fig:ToyScenarios}(f) shows a remarkable gain exceeding 188\% for C-SR with respect to DCF for all STAs except for STA$_{2}$, whose throughput is similar to the one under DCF since it is alone in its group. 

% --- Random
\subsection{Multiple Random Deployments}

\begin{figure}
    \centering
    \includegraphics[scale=0.6]{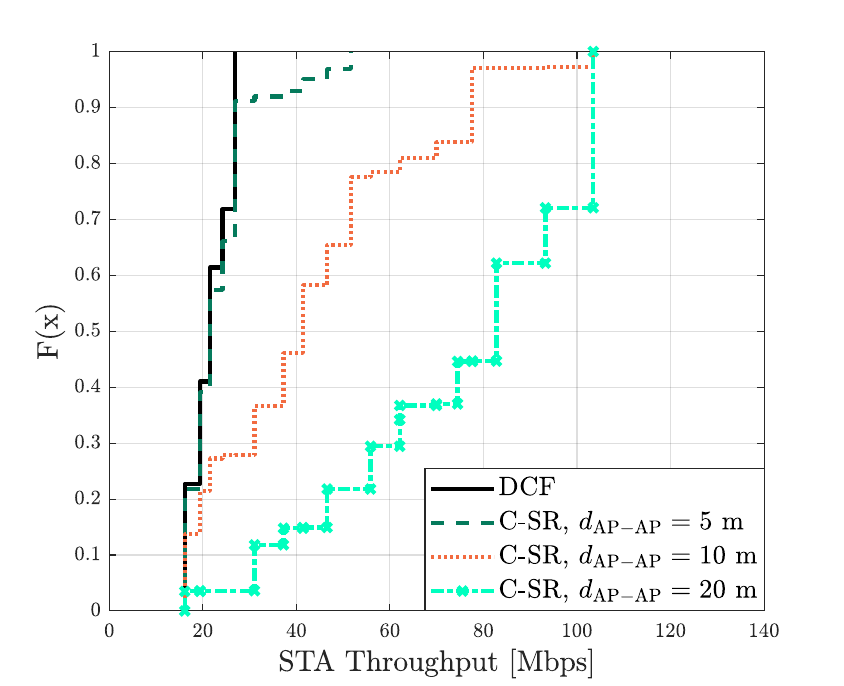}
    \caption{CDF of the STA throughput of DCF and C-SR, for $d_{\rm {AP-AP}} =~\{5,10,20\}$ meters and 10 STAs per AP.}
    \label{Fig:CDF}
\end{figure}

We also evaluate $1000$ random deployment realizations for $d_{\rm {AP-AP}} \in ~\{5, 10, 20\}$ meters, obtaining the throughput of each STA in all realizations. In each case, 10 STAs are deployed for each AP (for a total of 40 STAs) and placed $d_{\rm {AP-STA}} \in [1, 10]$ meters from their corresponding AP. Fig.~\ref{Fig:CDF} illustrates the Cumulative Distribution Function (CDF) for DCF (continuous-black line) and C-SR (dashed-green, dotted-red, and light-blue lines, for 5, 10, and 20 meters, respectively). We observe that C-SR outperforms DCF in all cases, achieving gains exceeding 73\%, 188\%, and 284\% for the 95$^{\rm th}$ percentile when $d_{\rm {AP-AP}} =~ 5$, $10$ and $20$ meters, respectively. As expected, when $d_{\rm {AP-AP}}$ decreases, the interference generated by concurrent AP transmissions makes it unfeasible to exploit spatial reuse opportunities, with C-SR performing similarly to DCF in most scenarios. On the contrary, C-SR transmissions become more profitable when $d_{\rm {AP-AP}}$ increases, thus for $d_{\rm {AP-AP}} = 10$ meters, only 1\% of the groups contain 4 APs, while for $d_{\rm {AP-AP}} = 20$ meters this proportion rises to 53\%.    

%\green{
% When comparing dotted red distribution in Fig.~\ref{Fig:CDF} (for $d_{\rm {AP-AP}} = 10$ meters and 10 STAs per AP) to the two examples in Fig.~\ref{Fig:ToyScenarios}(c) and Fig.~\ref{Fig:ToyScenarios}(f) (for $d_{\rm {AP-AP}} = 10$ meters and 1 STA per AP), we observe that increasing the number of STAs per AP corresponds to a less-than-proportional reduction in the per-STA throughput. Intuitively, the presence of more stations allows for a higher number of spatial reuse AP-STA compatible pairs, resulting in a more efficient network operation.%} 
%\green{Indeed, Fig.~\ref{Fig:ToyScenarios}(c) and Fig.~\ref{Fig:ToyScenarios}(f) show throughputs in the order of 150--800\,Mbps per STA with an average of about 400\,Mbps per STA, while the median in Fig.~\ref{Fig:CDF} for $d_{\rm {AP-AP}} = 10$ exceeds 60\,Mbps.} %\hl{I wrote it but not sure whether we should keep it or not.}

%a single group that contains all the four APs is the best option in the 20\% of the scenarios. 

% \begin{figure}[th]
%     \centering
%     \includegraphics[scale=0.6]{per_STA_thr_toy_scenario_DCF_cSR_PFT.pdf}
%     \caption{Per-STA throughput for DCF and C-SR in toy scenario of Fig.~\ref{Fig:scenario}, with 5 APs and 5 STAs. The C-SR in yellow has been obtained for $\phi_{i} = \frac{1}{3}$ while the red bar is C-SR with PFT$_{\rm {max}}$ and $\phi_{1}$, $\phi_{5}$, and $\phi_{7}$ are set to 0.4, 0.2, and 0.4, respectively.}
%     \label{Fig:thr_scenario_max_PFT}
% \end{figure}

% \begin{figure}[th]
%     \centering
%     \includegraphics[scale=0.6]{CDF_5APs_40STAs_agg_thr_DCF_C-SR_PFT.pdf}
%     \caption{CDF of the aggregate throughput for DCF and C-SR. Scenarios with 5 APs and 40 STAs randomly placed 1-4 meters away from their AP.}
%     \label{Fig:cdf_csma_cSR_PFT}
% \end{figure}

% \begin{figure}[th]
%     \centering
%     \includegraphics[scale=0.6]{bars_99th_average_1st_DCF_c-SR_PFT_10_20_30_40_STAs.pdf}
%     \caption{$99^{\rm th}$-percentile, average, and $1^{\rm st}$-percentile of the aggregate throughput for DCF and C-SR. Multiple scenarios with 5 APs and 10, 20, 30, and 40 randomly deployed STAs (1-4 meters away from their AP).}
%     \label{Fig:thr_99th_average_1th}
% \end{figure}

\begin{table}
    \caption{Simulation Parameters.}
    \scriptsize
    \label{tab:simulation_paramters}
    \begin{center}
        \begin{tabular}{|c|c|c|}
            \hline
           \textbf{Parameter}  & \textbf{Description} & \textbf{Value} \\
           \hline
            $d_{\rm {AP-AP}}$ & Minimum distance between APs [meters] & \{5, 10, 20\} \\
            \hline
            $d_{\rm {AP-STA}}$ & Distance between AP and STAs [meters] & [1, 10] \\
            \hline
            $B_{\rm {p}}$ & Break-point distance [meters]  & 10 \\
            \hline
            $W_{\rm {n}}$ & Number of  walls (1 wall every 10 meters) & \{0, 1, 2\}  \\
            \hline
            $\rm BW$ & Bandwidth [MHz] & 80 \\
            \hline
            $f_{\rm {c}}$ & Carrier frequency [GHz] & 6 \\
            \hline
            $N_{\rm {SS}}$ & Number of spatial streams & 2 \\
            \hline
            $T_{\rm {s}}$ & TXOP duration [ms] & 5 \\
            \hline
            $T_{\rm {c}}$  & Duration of a collision slot [$\mu$s] & 137 \\ 
            \hline
            $T_{\rm MAPC}$ & Coordination overheads [$\mu$s] & 286 \\
            \hline
            $T_{\rm BACK}$ & Block ACK [$\mu$s] & 100 \\
            \hline
            CW$_{\rm min}$  & Min contention window & 15 \\
            \hline
            CW$_{\rm max}$ & Max contention window & 1023 \\
            \hline
            $m$ & Number of backoff stages & 6 \\
            \hline
            $T_{\rm {e}}$ & Duration of an empty slot [$\mu$s] & 9 \\
            \hline
            $\gamma_{\textrm{CE}}$ & Capture effect threshold [dB] & 15 \\
            \hline
            EIRP & Effective Isotropic Radiated Power [dBm] & 23 \\
            \hline
            $L$ & Length of single data frame [bytes] & 1500 \\
            \hline
        \end{tabular}
    \end{center}
\end{table}

% ---------------------------------------------------
% ---------------------------------------------------
% ---------------------------------------------------
% ---------------------------------------------------
\section{Conclusions}\label{section:conclusions}

In this paper, we introduced a multi-AP coordination mechanism based on the formation of spatial reuse groups of AP-STA pairs, i.e., groups of devices that can transmit simultaneously. We leveraged Bianchi's model to analyze the C-SR mechanism and evaluate its potential performance under full-buffer conditions. Our study compared the throughput of C-SR against that of DCF and studied the trade-off between group size and interference depending on the locations of Wi-Fi devices. Our results highlighted the effectiveness of C-SR group formation and its potential for enhancing overall network performance, demonstrating a significant throughput improvement of up to $284\%$ in certain cases. This approach could be further extended by including adaptive power control.

%Group formation allows to optimize spatial reuse and maximize throughput in multi-AP WLANs. This approach can be extended to support adaptive power control to further optimize C-SR capabilities, to study alternative C-SR signaling protocols in terms of overheads, as well as to apply the group creation and analysis to other different scenarios.

While we only presented a set of illustrative results due to space constraints, our analysis enables the evaluation of different group creation algorithms and TXOP sharing strategies. It also allows to study the effect of tuning key 802.11 parameters such as the allowed transmission rates, channel contention settings, and setting different transmission durations (e.g., per group).
%opens the door to further study MAPC and C-SR achievable throughput in its multiple sides. On the one hand, as mentioned before, 

%\bcom{Explain future work may consider using the same model to study other scenarios, the use of power control to find better groups, etc. to test the impact of other signaling protocols (i.e., less/more overheads), etc.}

%In this paper, we have evaluated a multi-AP coordinated transmissions approach supporting C-SR on top of DCF. We also proposed a method to create groups of spatial reuse-compatible AP-STA pairs, and two different approaches for scheduling these groups. Results show a significant throughput improvement higher than  $15\%$ when C-SR is employed in some cases. Besides, the method of selecting the transmission probability of the spatial reuse groups, based on the maximization of the proportional fair throughput shows better per-STA results  and also in the aggregate throughput.   

% ---------------------------------------------------
% ---------------------------------------------------
% ---------------------------------------------------
% ---------------------------------------------------
\section{Acknowledgements}

D. Nunez and B. Bellalta were supported by grant Wi-XR PID2021-123995NB-I00 (MCIU/AEI/FEDER,UE), by MCIN/AEI under the MdM Program (CEX2021-001195-M), and by SGR 00955-2021 AGAUR. G. Geraci was in part supported by the Spanish Research Agency through grants PID2021-123999OB-I00, CEX2021-001195-M, and CNS2023-145384, by the UPF-Fractus Chair, and by the Spanish Ministry of Economic Affairs and Digital Transformation and the European Union NextGenerationEU. 

\bibliographystyle{IEEEtran}
\bibliography{main}

\end{document}